\documentclass[a4paper]{report}
\usepackage[utf8]{inputenc}
\usepackage[T1]{fontenc}
\usepackage{RJournal}
\usepackage{amsmath,amssymb,array}
\usepackage{booktabs}

\begin{document}

%% do not edit, for illustration only
\sectionhead{Contributed research article}
\volume{XX}
\volnumber{YY}
\year{20ZZ}
\month{AAAA}

\begin{article}
  % !TeX root = RJwrapper.tex
\title{SortedEffects: Sorted Causal Effects in R}
\author{by Shuowen Chen, Victor Chernozhukov, Iv{\'a}n Fern{\'a}ndez-Val and Ye Luo}

\maketitle

\abstract{
\citet{sorted:2018} proposed the sorted effect method for nonlinear regression models. This method consists of reporting percentiles of the partial effects, the sorted effects, in addition to the average effect commonly used to summarize the heterogeneity in the partial effects. They also propose to use the sorted effects to carry out classification analysis where the observational units are classified as most and least affected if their partial effect are above or below some tail sorted effects.  The R package \CRANpkg{SortedEffects} implements the estimation and inference methods therein and provides tools to visualize the results. This vignette serves as an introduction to the package and displays basic functionality of the functions within.
}
%%%%%%%%%%%%%%%%%%%%%%%%%%%%%%%%%%%%%%%%%%%%%%%%%%%%%%%%%%%%%%%%%%%%
%%%%%%%%%%%%%%%%%%%%%%%%%%%%%%%%%%%%%%%%%%%%%%%%%%%%%%%%%%%%%%%%%%%%
\section{The Sorted Effects Method}
Many empirical questions in econometrics, machine learning and statistics boil down to studying how changes in a key variable $T$ affect an outcome variable of interest $Y$, holding fixed some control variables $W$. Such effects are called predictive partial effects (PEs), or treatment or structural partial effects when they have a causal or structural interpretation. Depending on the context, researchers often work with models that feature nonlinearity in the key variable of interest or nonlinearity in parameters. The first kind of models includes mean and quantile regressions in which $T$ is interacted with $W$, while the second type includes generalized linear models such as logit and probit. The methods implemented in \CRANpkg{SortedEffects} are designed to estimate and make inference on PEs in nonlinear models.

In nonlinear models the PEs vary with the underlying control variables. Consider the probit model as an example. The conditional probability of $\{Y=1\}$ is
$$\Pr(Y = 1 \mid T=t,W=w) =\Phi(t\beta+w'\gamma),$$
where $\Phi$ denotes CDF of standard normal distribution. Then, the PE of changing $T$ on $Y$, holding $W$ fixed at $w$, is
$$\Phi(\beta+w'\gamma)-\Phi(w'\gamma)$$
when $T$ is binary with values $0$ and $1$, or
$$\phi(t\beta+w'\gamma)\beta, \quad \phi(u) = \partial \Phi(u)/ \partial u,$$
when $T$ is continuous. Since $W$ typically differ among observational units, the PEs are heterogeneous.

More generally, suppose we have a regression function $g(X)$ corresponding to some characteristic of $Y$ conditional on the covariates $X:=[T, W]$. Let $\Delta(x)$ denote the PE of $T$ on $Y$, holding $W$ fixed. Then,
$$\Delta(x)=g(t_{1}, w)-g(t_{0}, w)$$
if $T$ is discrete and takes values $t_0$ and $t_1$, or
$$\Delta(x)=\partial g(t, w)/ \partial t$$
if $T$ is continuous and $t \mapsto g(t,w)$ is differentiable. As $\Delta(x)$ is a function of $x$, the PE of an observational unit, $\Delta(X)$, is a random variable with a distribution induced by the distribution of $X$. A popular statistic to summarize the heterogeneity of $\Delta(X)$ is the average partial effect (APE):
$$E[\Delta(X)]=\int\Delta(x)d\mu(x),$$
where $\mu$ denotes the distribution of $X$ in the population of interest. However, the APE might provide an incomplete summary of $\Delta(X)$ as it neglects all the heterogeneity by design.

\citet{sorted:2018} proposed the sorted partial effect (SPE) method to provide a more complete summary of $\Delta(X)$. This method consists on reporting the entire set of PEs, sorted in increasing order and indexed by a user-specified ranking in the population of interest. More specifically, the SPEs are defined as percentiles of the PE in the population of interest, that is
$$\Delta^{*}_{\mu}(u)=u^{th}-\text{quantile of } \Delta(X), \quad X \sim \mu.$$
The SPEs are also useful to conduct classification analysis (CA). This analysis consists of two steps. First, classify the observational units with PEs above or below some thresholds defined by tail SPEs  in the most or least affected groups. Second, report and compare the moments or distributions of the outcome and covariates in the two groups.

To apply the methods in practice, we need to replace $\Delta$ and $\mu$ by sample analogs $\hat \Delta$ and $\hat \mu$, and quantify the sampling uncertainty. \citet{sorted:2018} derived the theoretical underpinnings of the resulting empirical SPEs. We refer the interested reader to that paper for details.

We expect that the method be helpful to a large audience. For example, medical researchers are often interested in estimating the treatment effect of a drug. Treated units can experience different effects due to individual characteristics such as age and health status. The \CRANpkg{SortedEffects} package thus provides a way to visualize the heterogeneity in the effect. Researchers can also define the most and least affected groups and compare their characteristics.

The rest of the vignette proceeds as follows. We first introduce the main functions within the \CRANpkg{SortedEffect} package and use an application to racial-based discrimination in mortgage lending to illustrate the command options. Then we provide another application to gender wage gap using the Current Population Survey (CPS) data. The empirical results show that the SPE method is effective in uncovering heterogeneous effects and demographic differences between the most and least affected groups in both applications.
%%%%%%%%%%%%%%%%%%%%%%%%%%%%%%%%%%%%%%%%%%%%%%%%%%%%%%%%%%%%%%%%%%%%
%%%%%%%%%%%%%%%%%%%%%%%%%%%%%%%%%%%%%%%%%%%%%%%%%%%%%%%%%%%%%%%%%%%%
\section{The \CRANpkg{SortedEffects} package}
%%%%%%%%%%%%%%%%%%%%%%%%%%%%%%%%%%%%%%%%%%%%%%%%%%%%%%%%%%%%%%%%%%%%
\subsection{Functions in the Package}
The package contains three main commands: \code{spe}, \code{ca} and \code{subpop}. The package adds three methods to the generic \code{plot()} (\code{plot.spe}, \code{plot.ca} and \code{plot.subpop}) and \code{summary()} (\code{summary.spe}, \code{summary.ca} and \code{summary.subpop}). In practice, users only need to type \code{plot} and \code{summary} since the generic commands will automatically dispatch the appropriate method. In this section we explain the options in each function respectively. Lastly we briefly explain the bootstrap procedures for inference and bias correction. We provide the mathematical expressions  and an application to racial discrimation in mortgage lending to facilitate the understanding of the options of the commands.
%%%%%%%%%%%%%%%%%%%%%%%%%%%%%%%%%%%%%%%%%%%%%%%%%%%%%%%%%%%%%%%%%%%%
\subsection{Sorted Partial Effects Function}
The command \code{spe} provides estimation and inference methods for the SPE and APE. The general syntax is:
\begin{example}
spe(fm, data, method = c("ols", "logit", "probit", "QR"), 
    var_type = c("binary", "continuous", "categorical"), var, compare, subgroup = NULL, 
    samp_weight = NULL, us = c(1:9)/10, alpha = 0.1, taus = c(5:95)/100, 
    b = 500, parallel = FALSE, ncores = detectCores(), seed = 1, bc = TRUE,
    boot_type = c("nonpar", "weighted"))
\end{example}
The option \code{fm} stores the user-specified regression formula to estimate the PEs, which assigns the outcome $Y$ on the left-hand-side of a \code{$\sim$} operator, and the covariates $X$ on the right-hand-side, separated by \code{+} operators. The option \code{data} specifies the data set that contains the variables for the analysis. The user needs to specify the variable of interest $T$ in \code{var} as a string. The package accommodates four regression methods to estimate the PE, $\Delta(x)$: OLS (default), probit (\code{"probit"}), logit (\code{"logit"}) and quantile regression (\code{"QR"}).\footnote{We use the \CRANpkg{quantreg} package to conduct quantile regression \citep{quantreg}.} If \code{"QR"} is called, then the user needs to further specify the quantile indexes with \code{taus}. The default is \code{taus = c(5:95)/100}.

The option \code{samp\_weight} allows the user to input sampling weights. When \code{samp\_weight = NULL}, the default, the package automatically uses a vector of ones, i.e. no sampling weights are used.

The package provides three options for the variable of interest \code{var}:
\begin{enumerate}
	\item \code{var\_type = "binary"} if \code{var} is a binary variable such as a black or female indicator.
	\item \code{var\_type = "categorical"} if \code{var} is a categorical (factor) variable with more than 2 levels. One example is means of transportation with labels \code{"bus"}, \code{"train"} or \code{"bike"}. In this case, the user needs to further specify the two labels to be compared with the option \code{compare}. If the user is interested in the effect of changing from bus to bike, then \code{compare = c("bus", "bike")}. If the data only have levels, say \code{"1", "2"} and \code{"3"}, the users need to input levels instead.
	\item \code{var\_type = "continuous"} if \code{var} is a continuous variable. The package obtains the PEs using a  central difference numerical derivative of the form:
	$$f'(x)=\lim_{h\rightarrow 0}\frac{f(x+h)-f(x-h)}{2h},$$
	where $h$ is set to be \code{1e-7}. We avoid the symbolic derivative because it cannot correctly interpret terms involving \code{I()} for transformations of the variables such as powers, and can therefore cause erroneous estimation. The code of this part is inspired by Thomas Leeper's vignette on the \CRANpkg{margins} package \citep{margins}.
\end{enumerate}

The option \code{subgroup} allows the user to specify the population of interest. The default is \code{NULL}, which corresponds to the entire population. If the user is interested in subpopulations, say households whose \code{income} is lower than $10,000$ dollars, and the data in use is called \code{Data}, then the user can set \code{subgroup = Data[, "income"] < 10000}. When $T$ is a binary treatment indicator, the user can specify that the population of interest is the treated population with  \code{subgroup = Data[, var] == 1}. Note that users cannot input subgroup directly to the \code{data} option or using the \code{subset} option because the SPE methods use the whole sample to estimate the PEs. The option \code{subgroup} only specifies the population for the estimation of $\mu$, the distribution of $X$.

The option \code{us} specifies the set of quantile indexes corresponding to the estimated SPE to be reported. The mathematical definition of each  empirical SPE is
$$\widehat{\Delta}^{*}_{\widehat{\mu}}(u) = u^{th}-\text{quantile of } \widehat{\Delta}(X), \quad X \sim \widehat{\mu}.$$

The empirical SPE function
$$\{u\mapsto \widehat{\Delta}^{*}_{\widehat{\mu}}(u): u\in\mathcal{U}\}, \quad \mathcal{U}\subset[0,1]$$
then outputs the SPEs with quantile indexes in the set $\mathcal{U}$. The option \code{us} specifies $\mathcal{U}$. For example, \code{us = c(0.25, 0.5, 0.75)} specifies to report the SPEs corresponding to the three quartiles.

The option \code{alpha} specifies the significance level of the confidence bands. The default is \code{alpha = 0.1}, i.e. $90\%$ confidence level. The option \code{b} specifies the number of bootstrap repetitions. The default is \code{b = 500}.

The package supports two types of bootstrap: nonparametric (\code{boot\_type = "nonpar"}) and weighted with standard exponential weights (\code{boot\_type = "weighted"}). The user can fix the random seed for bootstrap simulation with the option \code{seed} and decide whether or not to bias-correct the estimates with the option \code{bc}. The default is \code{bc = TRUE}. Bootstrap and bias-correction will be discussed in section 2.5. The package features parallel computing, which is convenient to speed up the bootstrap. The user can use the option \code{parallel} to turn on or off parallel computing. If \code{parallel = TRUE}, the option \code{ncores} allows the user to specify the number of cores in the parallel computing. The default is \code{ncores = detectCores()}, where \code{detectCores()} is a command of the package \CRANpkg{parallel}.

The output of \code{spe} is a list containing four components: \code{spe}, \code{ape}, \code{us} and \code{alpha}. As the names indicate, \code{spe} stores the results for the SPE, and \code{ape} stores the results for the APE. Each component is a list with four elements: the estimates, lower and upper bounds of the uniform confidence bands, and bootstrapped standard errors obtained as rescaled interquartile ranges; see  \citet{sorted:2018}. The other two components respectively store the percentile indices of the SPE and significance level of the confidence bands, which are used in the commands \code{plot.spe} and \code{summary.spe}.

\code{plot.spe} plots the result of \code{spe} in one graph that includes the SPEs, APE and their corresponding confidence bands. Its general syntax is
\begin{example}
plot.spe(object, ylim = NULL, main = NULL, sub = NULL, xlab = "Percentile Index", 
         ylab = "Sorted Effects", ...)
\end{example}
where \rt{object} is the output of \code{spe}. The range of the x-axis is fixed to be the range of user-specified quantile index \code{us}. The options \code{ylim}, \code{xlab} and \code{ylab} respectively denote range of the y-axis and labels of the two axes. The options \code{main} and \code{sub} allow users to specify the main and sub titles of the plot. The option \code{...} is an argument of the generic \code{plot} command that allows for further graphic parameters.

The syntax of \rt{summary.spe} is as follows
\begin{example}
summary.spe(object, result = c("sorted", "average"), ...)
\end{example}
If \code{result = "sorted"}, the method provides a table that contain the SPE percentile indexes, estimates, bootstrap standard errors, pointwise and uniform confidence bands. If \code{result = "average"}, the methods tabulates APE estimate, bootstrap standard error and confidence interval.

We illustrate the usage of the command with an empirical application to racial discrimination in mortgage lending. We use data on mortgage applications in Boston from 1990 (\citet{data}). The Federal Reserve Bank of Boston collected these data in relation to the Home Mortgage Disclosure Act (HMDA), which was passed to monitor minority access to the mortgage market. To retrieve the data from the package, issue the command
\begin{example}
data("mortgage")
\end{example}
The outcome variable, $Y$, is \code{deny}, a binary indicator for mortgage denial. The key variable of interest, $T$, is \code{black}, a binary indicator for the applicant being black, while the control variables, $W$, are financial and demographical characteristics that might affect the  mortgage decision of the bank. These characteristics include the debt-to-income ratio  (\code{p\_irat}), expenses-to-income ratio (\code{hse\_inc}), bad consumer credit (\code{ccred}), bad mortgage credit (\code{mcred}), credit problems (\code{pubrec}), denied mortgage insurance (\code{denpmi}), medium loan-to-house value (\code{ltv\_med}), high loan-to-house value (\code{ltv\_high}), self employed (\code{selfemp}), single (\code{single}), and high school graduate (\code{hischl}).  The regression formula for the estimation of the PEs is specified as:
\begin{example}
fm <- deny ~ black + p_irat + hse_inc + ccred + mcred + pubrec + ltv_med + ltv_high + 
      denpmi + selfemp + single + hischl
\end{example}
We invoke the \code{spe} command to calculate the bias-corrected estimates of the SPE at the quantile indexes $\{0.02, 0.03, \ldots, 0.98 \}$ for the entire population using a logit model.
\begin{example}
test <- spe(fm = fm, data = mortgage, var = "black", method = "logit", us = c(2:98)/100, 
            b = 500, bc = TRUE)
\end{example}
The output \code{test} includes the estimates and confidence bands for the APE and SPE in the entire population. We use \code{plot} to visualize the results.
\begin{example}
plot(x = test, ylim = c(0, 0.25), ylab = "Change in Probability", 
     main = "APE and SPE of Being Black on the Prob of Mortgage Denial",
     sub = "Logit Model")
\end{example}
\begin{figure}[ht]
	\centering
	\includegraphics[scale=0.5]{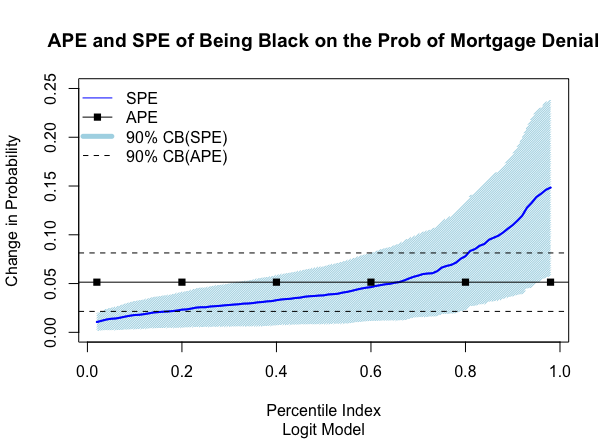}
	\caption{Output of \rt{plot.spe} function}\label{fig1}
\end{figure}
The result in Figure \ref{fig1} shows significant heterogeneity in the SPEs, with the PEs ranging from 0 to 14\%. The APE misses this heterogeneity, and therefore provides an incomplete picture of the effects.

We can also tabulate the result using \code{summary}. First, we apply the command to the APE and display the results in Table \ref{table1}.
\begin{example}
summary(test, result = "average")
\end{example}
\begin{table}[ht]
	\centering
	\caption{Bias-corrected estimates and bands of APE}\label{table1}
	\begin{tabular}{rrrrr}
		\hline
		& Est & SE & 90\% LB & 90\% UB \\
		\hline
		APE & 0.051 & 0.019 & 0.021 & 0.081 \\
		\hline
	\end{tabular}
\end{table}
Next, we obtain the results for the SPE. To save space we only show the first 15 rows of \code{summary(test)} in Table \ref{table2}. The columns labelled as PLB and PUB correspond to the lower and upper pointwise confidence bands, whereas the columns labelled as ULB and UUB correspond to their uniform counterparts.
\begin{example}
summary(test)
\end{example}
\begin{table}[ht]
	\centering
	\caption{Bias-corrected estimates and bands of SPE}\label{table2}
	\begin{tabular}{rrrrrrr}
		\hline
		& Est & SE & 90\% PLB & 90\% PUB & 90\% ULB & 90\% UUB \\
		\hline
		0.02 & 0.011 & 0.005 & 0.003 & 0.018 & 0.001 & 0.020 \\
		0.03 & 0.012 & 0.005 & 0.004 & 0.020 & 0.002 & 0.022 \\
		0.04 & 0.013 & 0.006 & 0.004 & 0.022 & 0.002 & 0.024 \\
		0.05 & 0.014 & 0.006 & 0.004 & 0.023 & 0.002 & 0.025 \\
		0.06 & 0.014 & 0.006 & 0.004 & 0.024 & 0.002 & 0.026 \\
		0.07 & 0.015 & 0.007 & 0.004 & 0.026 & 0.002 & 0.028 \\
		0.08 & 0.016 & 0.007 & 0.005 & 0.027 & 0.003 & 0.029 \\
		0.09 & 0.017 & 0.007 & 0.005 & 0.029 & 0.003 & 0.031 \\
		0.1 & 0.018 & 0.007 & 0.006 & 0.030 & 0.003 & 0.032 \\
		0.11 & 0.018 & 0.008 & 0.005 & 0.031 & 0.003 & 0.033 \\
		0.12 & 0.019 & 0.008 & 0.006 & 0.031 & 0.004 & 0.034 \\
		0.13 & 0.019 & 0.008 & 0.006 & 0.032 & 0.004 & 0.035 \\
		0.14 & 0.020 & 0.008 & 0.007 & 0.034 & 0.004 & 0.036 \\
		0.15 & 0.021 & 0.008 & 0.007 & 0.034 & 0.004 & 0.037 \\
		0.16 & 0.021 & 0.009 & 0.007 & 0.035 & 0.004 & 0.038 \\
		\hline
	\end{tabular}
\end{table}
%%%%%%%%%%%%%%%%%%%%%%%%%%%%%%%%%%%%%%%%%%%%%%%%%%%%%%%%%%%%%%%%%%%%
\subsection{Classification Analysis Function}
The command \code{ca} provides estimation and inference methods for the CA. The general syntax is:
\begin{example}
ca(fm, data, method = c("ols", "logit", "probit", "QR"),
   var_type = c("binary", "continuous", "categorical"), var,
   compare, subgroup = NULL, samp_weight = NULL, taus = c(5:95)/100,
   u = 0.1, interest = c("moment", "dist"),
   t = c(1, 1, rep(0, dim(data)[2] - 2)), cl = c("both", "diff"),
   cat = NULL, alpha = 0.1, b = 500, parallel = FALSE,
   ncores = detectCores(), seed = 1, bc = TRUE,
   range_cb = c(1:99)/100, boot_type = c("nonpar", "weighted"))
\end{example}
The first step in the CA is to classify the observational units in most and least affected groups based on some tail SPEs. The option \code{u} specifies the quantile index of the tail SPEs. Thus, the $u$-least affected group includes the observational units with $\widehat{\Delta}(X)<\widehat{\Delta}^{*}_{\widehat{\mu}}(u)$ and the $u$-most affected group the units with $\widehat{\Delta}(X)>\widehat{\Delta}^{*}_{\widehat{\mu}}(1-u)$. The default is \code{u = 0.1} to obtain the 10\% least and most affected groups. The option \code{subgroup} specifies the population of interest and has the same syntax as in the \code{spe} command.

Let $\widehat{\Lambda}^{-u}_{\widehat{\Delta}, \widehat{\mu}}(t)$ and $\widehat{\Lambda}^{+u}_{\widehat{\Delta}, \widehat{\mu}}(t)$ denote the  objects of interest in the least and most affected groups for the CA. Define
$$\widehat{\Lambda}^{u}_{\widehat{\Delta}, \widehat{\mu}}(t) := [\widehat{\Lambda}^{-u}_{\widehat{\Delta}, \widehat{\mu}}(t), \widehat{\Lambda}^{+u}_{\widehat{\Delta}, \widehat{\mu}}(t)].$$ These objects are indexed by the vector $t$, which specifies the variables of interest among the outcome and covariates. The option \code{t} is a vector that specifies $t$. Suppose the data has 5 variables (\code{"a", "b", "c", "d", "e"}) and we are interested in \code{"a"} and \code{"c"}, then we can either set \code{t = c("a", "c")} directly or \code{t = c(1, 0, 1, 0, 0)}. The second approach requires the user to know the order of the variables in the data set, which can be found with the command \code{View}.

Let $Z$ denote the set of variables of interest. The package provides two types of objects of interest. If \code{interest = "moment"}, then $\widehat{\Lambda}^{u}_{\widehat{\Delta}, \widehat{\mu}}(t)$ include the means of the variables in $Z$ for the least and most affected groups. If \code{interest = "dist"}, then $\widehat{\Lambda}^{u}_{\widehat{\Delta}, \widehat{\mu}}(t)$ includes the distributions of the variables in $Z$ for the least and most affected groups.

If \code{interest = "moment"}, \code{ca} estimates and makes inference on features of the chosen variables of interest in the least and most affected groups. These features are specified with the option \code{cl}. For example, if \code{cl = "both"}, the command estimates the mean of the variables in the two affected groups. On the other hand, \code{cl = "diff"} estimates the differences of the  means of the variables between the two groups.

If \code{interest = "dist"}, the option \code{range\_cb} specifies the region of interest for the domain of the distribution\footnote{Note that \code{cl} doesn't have any bearing when \code{interest = "dist"}.}. For example, if the variable of interest \code{x} is discrete, we can specify the region of interest as the support of \code{x} with \code{range\_cb = NULL}. If \code{x} is continuous, we can specify \code{range\_cb = c(1:99)/100} if we are interested in the percentiles with indexes $\{0.01, 0.03, \ldots, 0.99\}$. The default is \code{range\_cb = c(1:99)/100}. The choice \code{range\_cb = NULL} shuts down this feature by settimg the region of interest as all the distint values of the variable.

The output of \code{ca} depends on the choice of \code{interest}. For \code{interest = "moment"}, the output is a list containing the estimates of $\widehat{\Lambda}^{u}_{\widehat{\Delta}, \widehat{\mu}}(t)$, bootstrapped standard errors, pointwise and adjusted p-values. The null hypothesis for the p-values is that the estimated coefficient is zero. The p-values are adjusted for multiplicity to account for joint testing for all variables. In addition, users can adjust the pointwise p-values to account for joint testing for all simulataneous tests of categories within a factor. For example, if the variables of interest include a marital status factor \code{"ms"} with labels \code{("nevermarried", "married", "divorced", "separated", "widowed")}, then users could consider adjusting the pointwise p-values within this factor. To illustrate how to define the option \code{cat}, suppose we have selected specified 3 variables of interest: \code{t = c("a", "b", "c")}. Without loss of generality, assume \code{"a"} is not a factor, while \code{"b"} and \code{"c"} are two factors. Then we need to specify \code{cat} as \code{cat = c("b", "c")}. If \code{cat = NULL}, we report the unadjusted pointwise p-values. If \code{interest = "dist"}, the output is a list containing the rearranged estimates, upper confidence bands and lower confidence bands for the variables of interest in both groups.

When \code{interest = "moment"}, the user can use method \code{summary.ca} to tabulate the output. The general syntax is
\begin{example}
summary.ca(object, ...)
\end{example}
If \code{cl = "both"},  the p-values are omitted from the table. When \code{interest = "dist"}, users can plot the output for better visualization. The general syntax is
\begin{example}
plot.ca(object, var, main = NULL, sub = NULL, xlab = NULL, ylab = NULL, ...)
\end{example}
The user needs to input the variable for plotting with the option \code{var}. Note that the variable must be one of the variables specified in \code{t}.

Returning to the mortgage denial example, we classify the 10\% least and most affected applicants and compare their characteristics. The variables of interest include \code{deny}, \code{black} and all the controls. We first specify \code{t} to reflect the choice of variables of interest\footnote{Alternatively, we can use \code{View(mortgage)} to locate the variables and set \code{t <- c(rep(1, 4), 0, rep(1, 7), 0, 0, 1, 1)}.}
\begin{example}
t <- c("deny", "p_irat", "black", "hse_inc", "ccred", "mcred", "pubrec", "denpmi", 
       "selfemp", "single", "hischl", "ltv_med", "ltv_high")
\end{example}
Then we invoke the \rt{ca} command and summarize the result.
\begin{example}
CA <- ca(fm = fm, data = mortgage, var = "black", method = "logit", cl = "both", 
         t = t, b = 500, bc = TRUE)
summary(CA)
\end{example}
\begin{table}[ht]
	\centering
	\caption{Bias-corrected mean characteristics of the 10\% most and least affected groups}\label{table3}
	\begin{tabular}{rrrrr}
		\hline
		& Most & SE & Least & SE \\
		\hline
		deny & 0.45 & 0.03 & 0.09 & 0.04 \\
		p\_irat & 0.39 & 0.01 & 0.25 & 0.02 \\
		black & 0.38 & 0.03 & 0.06 & 0.02 \\
		hse\_inc & 0.28 & 0.01 & 0.21 & 0.02 \\
		ccred & 4.80 & 0.26 & 1.28 & 0.09 \\
		mcred & 2.01 & 0.06 & 1.36 & 0.10 \\
		pubrec & 0.46 & 0.05 & 0.05 & 0.02 \\
		denpmi & 0.01 & 0.01 & 0.04 & 0.03 \\
		selfemp & 0.17 & 0.04 & 0.04 & 0.03 \\
		single & 0.61 & 0.06 & 0.09 & 0.07 \\
		hischl & 0.93 & 0.03 & 1.00 & 0.01 \\
		ltv\_med & 0.59 & 0.06 & 0.05 & 0.04 \\
		ltv\_high & 0.12 & 0.04 & 0.01 & 0.01 \\
		\hline
	\end{tabular}
\end{table}
Table \ref{table3} shows that the 10\% of the applicants most affected by the racial mortgage denial gap are more likely to have either of the following characteristics relative to the 10\% of the least affected applicants: mortgage denied, high debt-to-income ratio, black, high expense-to-income ratio, bad consumer or credit scores, credit problems, self employed, single, no high school diploma, and medium or high loan-to-income ratio.

Next we test if the differences in the characteristics between the two groups are statistically significant. To do so we set \code{cl = "diff"}, which means taking difference between the two groups. The full command is as follows
\begin{example}
CAdiff <- ca(fm = fm, data = mortgage, var = "black", t = t, method = "logit", 
             cl = "diff", b = 500, bc = TRUE)
summary(CAdiff)
\end{example}
\begin{table}[ht]
	\caption{Bias-corrected mean difference of characteristics of the 10\% most and least affected groups}\label{table4}
	\centering
	\begin{tabular}{rrrrr}
		\hline
		& Estimate & SE & JP-vals & P-vals \\
		\hline
		deny & 0.36 & 0.05 & 0.00 & 0.00 \\
		p\_irat & 0.14 & 0.02 & 0.00 & 0.00 \\
		black & 0.32 & 0.04 & 0.00 & 0.00 \\
		hse\_inc & 0.07 & 0.02 & 0.13 & 0.00 \\
		ccred & 3.52 & 0.28 & 0.00 & 0.00 \\
		mcred & 0.65 & 0.15 & 0.01 & 0.00 \\
		pubrec & 0.41 & 0.05 & 0.00 & 0.00 \\
		denpmi & -0.03 & 0.04 & 1.00 & 0.23 \\
		selfemp & 0.13 & 0.06 & 0.48 & 0.02 \\
		single & 0.53 & 0.10 & 0.00 & 0.00 \\
		hischl & -0.06 & 0.03 & 0.48 & 0.02 \\
		ltv\_med & 0.54 & 0.07 & 0.00 & 0.00 \\
		ltv\_high & 0.10 & 0.03 & 0.10 & 0.00 \\
		\hline
	\end{tabular}
\end{table}
Table \ref{table4} shows the results. The joint p-values account for the fact that we conduct simultaneous inference on 13 differences of variables. We employ the so-called ``single-step'' methods for controlling the family-wise error rate and obtain the p-values by bootstrap. We find that 8 differences are jointly statistically different from zero at the 5\% level and 9 at the 10\% level.

We also plot the distributions of monthly debt-to-income ratio (\code{p\_irat}) and monthly housing expenses-to-income ratio (\code{hse\_inc}) for both groups. Such plots are useful if the user wants to visualize if there is stochastic dominance between the two groups. To do so we use the \code{ca} command and change the \code{interest} to \code{"dist"}.
\begin{example}
t2 <- c("p_irat", "hse_inc")
CAdist <- ca(fm = fm, data = mortgage, var = "black", method = "logit", t = t2, 
             b = 500, interest = "dist")
plot(CAdist, var = "p_irat", ylab = "Prob", xlab = "Monthly Debt-to-Income Ratio", 
     sub = "logit model")
plot(CAdist, var = "hse_inc", ylab = "Prob", 
     xlab = "Monthly Housing Expenses-to-Income Ratio", sub = "logit model")
\end{example}
\begin{figure}[h]
	\centering
	\begin{minipage}[b]{0.495\textwidth}
		\includegraphics[width=\textwidth]{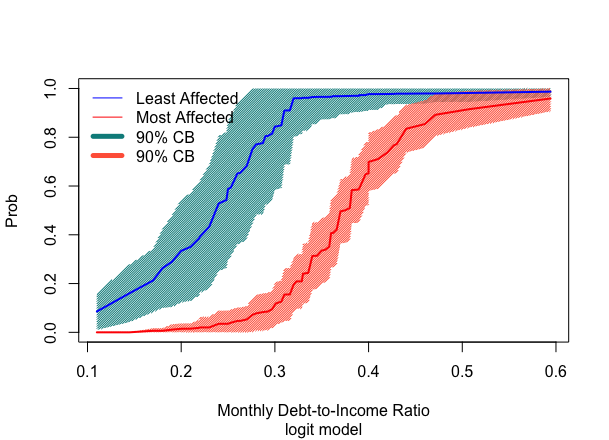}
	\end{minipage}
	\hfill
	\begin{minipage}[b]{0.495\textwidth}
		\includegraphics[width=\textwidth]{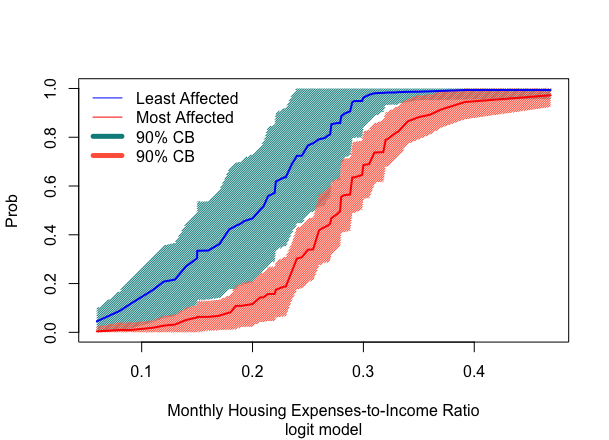}
	\end{minipage}
	\caption{Output of \rt{plot.ca} function}\label{fig2}
\end{figure}
Figure \ref{fig2} shows that for both variables the distribution in the most affected group first-order stochastically dominates the distribution in least affected group.
%%%%%%%%%%%%%%%%%%%%%%%%%%%%%%%%%%%%%%%%%%%%%%%%%%%%%%%%%%%%%%%%%%%%
\subsection{Subpopulations Function}
In addition to means and distributions, we can conduct inference on the sets of most and least affected units. Let $\mathcal{Z}$ be a compact subset of the support of the outcome and covariates. Define
$$\mathcal{M}^{-u}\equiv \{(x,y)\in\mathcal{Z}: \Delta(x)\leq\Delta^{*}_{\mu}(u)\}$$
as the set of the least affected units and
$$\mathcal{M}^{+u}\equiv \{(x,y)\in\mathcal{Z}: \Delta(x)\geq\Delta^{*}_{\mu}(1-u)\}$$
as the set of the most affected units. The command \code{subpop} provides estimation and inference methods for these sets. The general syntax is
\begin{example}
subpop(fm, data, method = c("ols", "logit", "probit", "QR"),
       var_type = c("binary", "continuous", "categorical"), var, compare, 
       subgroup = NULL, samp_weight = NULL, taus = c(5:95)/100, u = 0.1, alpha = 0.1, 
       b = 500, seed = 1, parallel = FALSE, ncores = detectCores(), 
       boot_type = c("nonpar", "weighted"))
\end{example}
No new option is introduced in the command. For theoretical details, we refer the reader to \citet{setinf}. 

The output of \code{subpop} is a list containing six components: \code{cs\_most}, \code{cs\_least}, \code{u}, \code{subgroup}, \code{most} and \code{least}. As the names indicate, \code{cs\_most} and \code{cs\_least} denote the confidence sets for the most and least affected groups. \code{u} stores the percentile index that defines the most and least affected groups. \code{subgroup} stores the indicators for the population of interest specified with the option \code{subgroup}. \code{most} and \code{least} store the estimates of the most and least affected units respectively. The first four components are used in \code{plot.subpop} and the last two components can be visualized with \code{summary.subpop}.

The general syntax of \code{summary.subpop} is
\begin{example}
summary.subpop(object, affected = c("most", "least"), vars = NULL, ...)
\end{example}
The option \code{object} is the output of \code{subpop}. The option \code{affected} allows users to tabulate either the most or the least affected units, and the option \code{vars} provides summary statistics for user-specified variables of interest. The summary statistics include the minimum, 1st quartile, median, mean, 3rd quaritle and maximum. The default is \code{NULL}, which produces summary statistics of all the variables.

\code{plot.subpop} plots 2-dimensional projections of the confidence sets for the most and least affected units with respect to two variables. The general syntax is
\begin{example}
plot.subpop(object, varx, vary, xlim = NULL, ylim = NULL, main = NULL, sub = NULL, 
            xlab = NULL, ylab = NULL, overlap = FALSE, ...)
\end{example}
The user needs to specify the two variables for the projection with \code{varx} and \code{vary}, and \code{object} should be specified as the output of \code{subpop}. The option \code{overlap} allows users to either keep or drop common observations in both confidence sets. The default is \code{overlap = FALSE}, which drops the observations.

\begin{figure}[ht]
	\centering
	\includegraphics[scale=0.5]{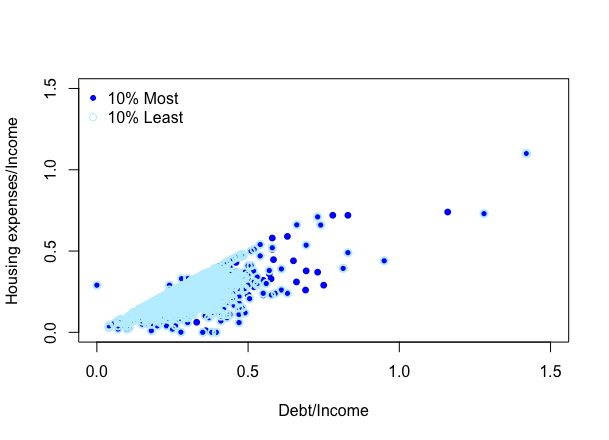}
	\caption{Projections of confidence sets of 10\% most and least affected applicants}\label{fig3}
\end{figure}

We estimate the 10\% most and least affected applicants in the mortgage application.
\begin{example}
set_b <- subpop(fm, data = mortgage, method = "logit", var = "black", u = 0.1, 
                alpha = 0.1, b = 500)
\end{example}
Using \code{summary}, we can estimate the most/least affected applicants and report summary statistics of the variables of interest in the most/least affected groups. Table \ref{table5} lists the estimated most affected applicants. For the purpose of illustration we only show the first ten rows and columns.
\begin{example}
groups <- summary(set_b, vars = c("p_irat", "hse_inc"))
most_affected <- groups$most_affected
\end{example}
\begin{table}[ht]
\centering
\caption{Applicants in 10\% most affected group}\label{table5}
\begin{tabular}{rrrrrrrrrrr}
\hline
& deny & p\_irat & black & hse\_inc & loan\_val & ccred & mcred & pubrec & denpmi & selfemp \\
\hline
1 & 1.00 & 0.46 & 0.00 & 0.27 & 0.84 & 5.00 & 2.00 & 0.00 & 0.00 & 0.00 \\
2 & 1.00 & 0.38 & 0.00 & 0.26 & 0.88 & 6.00 & 1.00 & 1.00 & 0.00 & 0.00 \\
3 & 0.00 & 0.40 & 0.00 & 0.34 & 0.80 & 2.00 & 2.00 & 0.00 & 0.00 & 1.00 \\
4 & 1.00 & 0.24 & 0.00 & 0.23 & 0.90 & 5.00 & 2.00 & 0.00 & 0.00 & 0.00 \\
5 & 0.00 & 0.38 & 0.00 & 0.25 & 0.80 & 6.00 & 2.00 & 1.00 & 0.00 & 0.00 \\
6 & 0.00 & 0.36 & 0.00 & 0.13 & 0.95 & 5.00 & 2.00 & 0.00 & 0.00 & 0.00 \\
7 & 0.00 & 0.35 & 0.00 & 0.27 & 0.90 & 6.00 & 2.00 & 0.00 & 0.00 & 0.00 \\
8 & 1.00 & 0.30 & 0.00 & 0.30 & 0.50 & 6.00 & 2.00 & 1.00 & 0.00 & 0.00 \\
9 & 1.00 & 0.37 & 0.00 & 0.23 & 0.80 & 6.00 & 2.00 & 1.00 & 0.00 & 0.00 \\
10 & 1.00 & 0.39 & 0.00 & 0.27 & 0.90 & 6.00 & 2.00 & 1.00 & 0.00 & 0.00 \\
\hline
\end{tabular}
\end{table}
We can also report summary statistics of the variables of interest in the most and least affected groups using the output of \code{summary}. Table \ref{table6} reports summary statistics for \code{p\_irat} and \code{hse\_inc} for the applicants in the most affected group.
\begin{example}
sum_stats_most <- groups$stats_most
\end{example}
\begin{table}[ht]
	\caption{Summary statistics of for the most affected group}\label{table6}
	\centering
	\begin{tabular}{rrr}
		\hline
		& p\_irat & hse\_inc \\
		\hline
		Min & 0.16 & 0.01 \\
		1st Quartile & 0.34 & 0.23 \\
		Median & 0.37 & 0.28 \\
		Mean & 0.39 & 0.28 \\
		3rd Quartile & 0.42 & 0.32 \\
		Max & 1.16 & 0.74 \\
		\hline
	\end{tabular}
\end{table}
We finally plot the projection of the confidence sets for the most and least affected applicants with respect to \code{p\_irat} and \code{hse\_inc}. Figure \ref{fig3} keeps the overlapped observations and shows that the most affected applicants tend to have higher levels of debt to income and expenses to income ratios.
\begin{example}
plot(set_b, varx = mortgage$p_irat, vary = mortgage$hse_inc, xlim = c(0, 1.5), 
     ylim = c(0, 1.5), xlab = "Debt/Income", ylab = "Housing expenses/Income", 
     overlap = TRUE)
\end{example}
%%%%%%%%%%%%%%%%%%%%%%%%%%%%%%%%%%%%%%%%%%%%%%%%%%%%%%%%%%%%%%%%%%%%
\subsection{Inference}
\citet{sorted:2018} derived the asymptotic distributions and bootstrap validity for the estimators of the SPE and classification analysis. The package uses bootstrap to compute standard errors and critical values for tests and confidence bands.

The package features nonparametric and weighted bootstrap. When \code{boot\_type = "nonpar"}, the package draws samples with replacement of the variables and \code{samp\_weight} and run all estimation commands weighted by \code{samp\_weight}. When \code{boot\_type = "weighted"}, the package draws weights from the standard exponential distribution and runs all estimation commands weighted by the product of these weights and \code{samp\_weight}. We use the \CRANpkg{boot} package \citep{boot}, which is flexible enough to accommodate both types.

\noindent \textbf{Inference on SPE } The $(1-\alpha)$-uniform confidence band of $\Delta^{*}_{\mu}(u)$ in $\mathcal{U}$ is
$$[\widehat{\Delta}^{*}_{\widehat{\mu}}-\widehat{t}_{1-\alpha}(\mathcal{U})\widehat{\Sigma}(u)^{1/2}/\sqrt{n}, \quad \widehat{\Delta}^{*}_{\widehat{\mu}}+\widehat{t}_{1-\alpha}(\mathcal{U})\widehat{\Sigma}(u)^{1/2}/\sqrt{n}],$$
where $\widehat{t}_{1-\alpha}(\mathcal{U})$ is a bootstrapped uniform critical value and $\widehat{\Sigma}(u)^{1/2}$ is a boostrapped standard error of $\widehat{\Delta}^{*}_{\widehat{\mu}}(u)$.\footnote{See Algorithm 2.1 in \citet{sorted:2018} for details.} To deal with the possibility that the end-point functions of the confidence band $u \mapsto \widehat{\Delta}^{*}_{\widehat{\mu}}\pm\widehat{t}_{1-\alpha}(\mathcal{U})\widehat{\Sigma}(u)^{1/2}/\sqrt{n}$ be nonincreasing, we monotonize these functions via rearrangement \citep{rearrangement}. The $(1-\alpha)$-pointwise confidence band of $\Delta^{*}_{\mu}(u)$ is obtained replacing $\widehat{t}_{1-\alpha}(\mathcal{U})$ by the $(1-\alpha/2)$-quantile of the standard normal distribution.

\noindent \textbf{Inference on CA } The joint $p$-value for the hypothesis $\Lambda^{+u}_{\Delta, \mu}(t) = \Lambda^{-u}_{\Delta, \mu}(t)$ for all $t\in\mathcal{T}$  is
$$\Pr\left(\hat{t}^{u}(\mathcal{T}) > \sup_{t\in\mathcal{T}} \frac{|c'\widehat{\Lambda}^{u}_{\widehat{\Delta}, \widehat{\mu}}(t)c|}{\sqrt{c'\hat{\Sigma}^{u}(t) c}}\right),$$
where $c=(-1,1)'$, $\hat{t}^{u}(\mathcal{T})$ is a bootstrap estimator of
$$
t^{u}(\mathcal{T}) = \sup_{t\in\mathcal{T}} \frac{|c'\widehat{\Lambda}^{u}_{\widehat{\Delta}, \widehat{\mu}}(t)c - c'\Lambda^{u}_{\Delta, \mu}(t)c|}{\sqrt{c'\Sigma^{u}(t) c}},
$$
and $\hat{\Sigma}^{u}(t)$ is a bootstrap estimator of $\Sigma^{u}(t)$, the asymptotic variance of $\widehat{\Lambda}^{u}_{\widehat{\Delta}, \widehat{\mu}}(t)$.\footnote{See Algorithm 2.2 in \citet{sorted:2018} for details.} The poitwise $p$-value for the hypothesis $\Lambda^{+u}_{\Delta, \mu}(t) = \Lambda^{-u}_{\Delta, \mu}(t)$ is obtained by setting $\mathcal{T} = \{t\}$.

\noindent \textbf{Inference on Sets of Least/Most Affected Units } The outer $(1-\alpha)$-confidence set for $\mathcal{M}^{-u}$ is
$$\mathcal{C}\mathcal{M}^{-u}(1-\alpha)=\{(x, y)\in\mathcal{Z}: \widehat{\Sigma}^{-1/2}(x, u)\sqrt{n}[\widehat{\Delta}(x)-\widehat{\Delta}^{*}_{\mu}(u)]\leq\hat{c}(1-\alpha)\},$$
where $\widehat{\Sigma}(x, u)$ is an estimator of the asymptotic variance of $\widehat{\Delta}(x)-\widehat{\Delta}^{*}_{\mu}(u)$.
Similarly the outer $(1-\alpha)$-confidence set for $\mathcal{M}^{+u}$ is
$$\mathcal{C}\mathcal{M}^{+u}(1-\alpha)=\{(x, y)\in\mathcal{Z}: \widehat{\Sigma}^{-1/2}(x, 1-u)\sqrt{n}[\widehat{\Delta}^{*}_{\mu}(1-u)-\widehat{\Delta}(x)]\leq\tilde{c}(1-\alpha)\}.$$
The critical value $\hat{c}(1-\alpha)$ is the $(1-\alpha)$-quantile of the statistic:
$$\tilde{V}^{*}_{\infty}=\sup_{\{x\in\mathcal{X}:\widehat{\Delta}(x)=\widehat{\Delta}^{*}_{\mu}(u)\}}\widehat{\Sigma}^{-1/2}(x, u)\sqrt{n}([\tilde{\Delta}(x)-\tilde{\Delta^{*}_{\mu}}(u)]-[\widehat{\Delta}(x)-\widehat{\Delta}^{*}_{\mu}(u)]).$$
while the critical value $\tilde{c}(1-\alpha)$ is the $(1-\alpha)$-quantile of the statistic:
$$\tilde{V}^{**}_{\infty}=\sup_{\{x\in\mathcal{X}:\widehat{\Delta}(x)=\widehat{\Delta}^{*}_{\mu}(1-u)\}}\widehat{\Sigma}^{-1/2}(x, 1-u)\sqrt{n}([\tilde{\Delta}(x)-\tilde{\Delta^{*}_{\mu}}(1-u)]-[\widehat{\Delta}(x)-\widehat{\Delta}^{*}_{\mu}(1-u)]).$$
To implement $\sup\{x\in\mathcal{X}: \widehat{\Delta}(x)=\widehat{\Delta}^{*}_{\mu}(u)\}$ in the code, we find the minimum of $|\widehat{\Delta}(x)-\widehat{\Delta}^{*}_{\widehat{\mu}}(u)|$ among all $x$'s.

\noindent \textbf{Bias-correction } Nonlinear estimators are prone to finite-sample bias, and bootstrap methods can estimate the bias up to some asymptotic order. To bias correct the SPE, replace $\widehat{\Delta}^{*}_{\widehat{\mu}}$ with $2\widehat{\Delta}^{*}_{\widehat{\mu}}-\overline{\Delta^{*}_{\mu}}$, where $\overline{\Delta^{*}_{\mu}}$ is the mean of bootstrap draws. Similarly, for CA we replace $\widehat{\Lambda}^{u}_{\widehat{\Delta}, \widehat{\mu}}(t)$ with $2\widehat{\Lambda}^{u}_{\widehat{\Delta}, \widehat{\mu}}(t) - \overline{\Lambda^{u}_{\Delta, \mu}}$, where $\overline{\Lambda^{u}_{\Delta, \mu}}$ is the mean of the bootstrap draws. Bias-corrected estimates and corresponding inference will be reported if \rt{bc = TRUE}, the default.
%%%%%%%%%%%%%%%%%%%%%%%%%%%%%%%%%%%%%%%%%%%%%%%%%%%%%%%%%%%%%%%%%%%%
%%%%%%%%%%%%%%%%%%%%%%%%%%%%%%%%%%%%%%%%%%%%%%%%%%%%%%%%%%%%%%%%%%%%
\section{Gender Wage Gap Application}
We  analyze the gender wage gap using data from the U.S. March Supplement of the Current Population Survey (CPS) in 2015. The gender wage gap measures the difference in wages between female and male workers with the same observable characteristics. The SPE method allows us to look for heterogeneity in the gender wage gap and to identify the characteristics of the most and least affected workers. To retrieve the data, issue the command
\begin{example}
data(wage2015)
\end{example}
The data contain the following variables: log hourly wages (\code{lnw}); a marital status factor \code{ms} with 5 categories \code{widowed, divorced, separated, nevermarried, married}; CPS sampling weights (\code{weight}); a indicator for female worker (\code{female}); an education attainment factor \code{educ} with 5 categories \code{lhs} (less than high school graduate), \code{hsg} (high school graduate), \code{sc} (some college), \code{cg} (college) and \code{ad} (advanced degree); a region factor \code{region} with 4 categories \code{mw} (midwest), \code{so} (south), \code{we} (west) and \code{ne} (northeast); potential work experience \code{exp1} computed  as $\max\{0, \text{age - years of educ} -7\}$; 4 powers of experience (\code{exp2, exp3, exp4}); an occupation factor \code{occ} with 5 categories \code{manager}, \code{service}, \code{sales}, \code{construction} and \code{production}, and an industry factor \code{ind} with 12 categories \code{minery}, \code{construction}, \code{manufacture}, \code{retail}, \code{transport}, \code{information}, \code{finance}, \code{professional}, \code{education}, \code{leisure}, \code{services} and \code{public}.

The CPS data contains sampling weights in the variable \code{weight}, so we will set \code{samp\_weight = wage2015\$weight}. Because women in general earn less than men, the PEs are predominately negative if we use the \code{female} indicator. To facilitate the interpretation of most and least affected groups we create an indicator called \code{male}, which assigns 0 to female workers instead.
\begin{example}
wage2015$male <- 1 - wage2015$female
\end{example}
We apply OLS regression to estimate the PEs using the following specification
\begin{example}
fmla1 <- lnw ~ male*(ms + (exp1 + exp2 + exp3 + exp4)*educ + occ + ind + region)
\end{example}
We first look at the SPE of the gender wage gap at the quantile indexes $\{0.02, 0.03, \ldots, 0.98\}$ in the population of women via the command \code{spe} and plot the result. We specify that the population of interest is female workers with \code{subgroup = wage2015[,"female"] == 1}.
\begin{example}
gap <- spe(fm = fmla1, data = wage2015, samp_weight = wage2015$weight, 
           var = "male", subgroup = wage2015[,"female"] == 1, boot_type = "weighted", 
           us = c(2:98)/100, b = 500, bc = FALSE)

plot(x = gap, main = "APE and SPE of Gender Wage Gap for Women", sub = "OLS Model", 
     xlab = "Percentile Index", ylab = "Gender Wage Gap", ylim = c(-0.1, 0.45))
\end{example}
\begin{figure}[ht]
	\centering
	\includegraphics[scale=0.5]{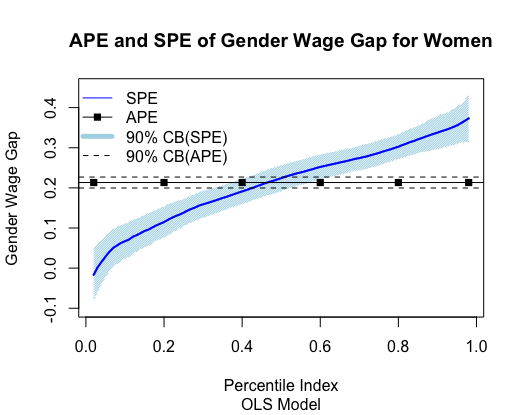}
	\caption{APE and SPE of the gender wage gap for women}\label{fig4}
\end{figure}
Figure \ref{fig4} shows large heterogeneity in the gender wage gap that is missed if we only report the APE.

\begin{figure}[ht]
	\centering
	\begin{minipage}[b]{0.495\textwidth}
		\includegraphics[width=\textwidth]{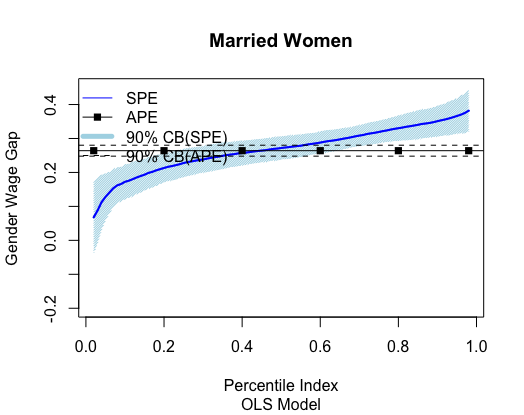}
	\end{minipage}
	\hfill
	\begin{minipage}[b]{0.495\textwidth}
		\includegraphics[width=\textwidth]{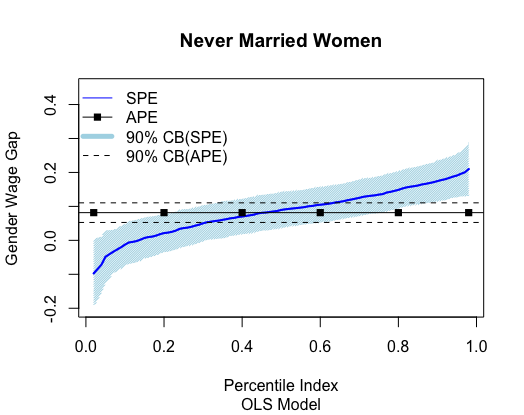}
	\end{minipage}
	\caption{SPE and APE of gender wage gap for married and never married women}\label{fig5}
\end{figure}

We also compare the SPE across subsets of women defined by marital status. We implement this by changing the \code{subgroup} options as follows
\begin{example}
fem_mar <- wage2015[, "female"] == 1 & wage2015[, "ms"] == "married"
fem_nev <- wage2015[, "female"] == 1 & wage2015[, "ms"] == "nevermarried"

gap_mar <- spe(fm = fmla1, data = wage2015, samp_weight = wage2015$weight, 
               var = "male",  subgroup = fem_mar, us = c(2:98)/100, b = 500, 
               bc = FALSE, boot_type = "weighted")

gap_nev <- spe(fm = fmla1, data = wage2015, samp_weight = wage2015$weight,
               var = "male", subgroup = fem_nev, us = c(2:98)/100, b = 500,
               bc = FALSE, boot_type = "weighted")

plot(x = gap_mar, main = "Married Women", sub = "OLS Model", xlab = "Percentile Index", 
     ylab = "Gender Wage Gap", ylim = c(-0.2, 0.45))

plot(x = gap_nev, main = "Never Married Women", sub = "OLS Model", 
     xlab = "Percentile Index",  ylab = "Gender Wage Gap", ylim = c(-0.2, 0.45))
\end{example}
Figure \ref{fig5} shows the results for the two subpopulations. Here we find large heterogeneity not only between married and never married women, but also within these more narrowly defined subpopulations.
Now we compare the differences in characteristics of the 5\% most and least affected women using weighted bootstrap with $500$ repetitions. We pick the following variables
\begin{example}
tw <- c("lnw", "female", "ms", "educ", "region", "exp1", "occ", "ind")
\end{example}
Since many variables arecategory indicators of the same factor, we can specify \code{cat} as follows to get categorical p-values.
\begin{example}
cat <- c("ms", "educ", "region", "occ", "ind")
\end{example}
Then we issue the \code{ca} command and tabulate the mean characteristics of the two groups
\begin{example}
Char <- ca(fm = fmla1, data = wage2015, samp_weight = wage2015$weight, var = "male", 
           t = tw, cl = "both", b = 500, subgroup = wage2015[,"female"] == 1, 
           boot_type = "weighted", bc = FALSE, u = 0.05)
\end{example}
Table \ref{table7} reports the estimates and standard errors obtained by weighted bootstrap with 500 replications. We find that, compared to the 5\% least affected women, the 5\% most affected women are much more likely to be married, much less likely to be never married, less likely to have an advanced degree, live in the South, don't live in Northeast and West, possess more potential experience, are more likely to have sales and non-managerial occupations, and work more often in manufacture, retail, transport and finance, and less often in education and leisure industries.
\begin{example}
summary(Char)
\end{example}
\begin{table}[h]
\centering
\caption{Mean characteristics of 5\% most and least affected women}\label{table7}
\begin{tabular}{lrrrr}
\hline
& Most & SE & Least & SE \\
\hline
lnw & 2.97 & 0.10 & 3.02 & 0.06 \\
female & 1.00 & 0.00 & 1.00 & 0.00 \\
exp1 & 26.58 & 2.01 & 8.47 & 2.54 \\
occ\_manager & 0.21 & 0.13 & 0.77 & 0.08 \\
occ\_service & 0.04 & 0.03 & 0.12 & 0.06 \\
occ\_sales & 0.56 & 0.14 & 0.10 & 0.05 \\
occ\_construction & 0.00 & 0.01 & 0.01 & 0.01 \\
occ\_production & 0.19 & 0.08 & 0.01 & 0.01 \\
ind\_minery & 0.00 & 0.02 & 0.00 & 0.00 \\
ind\_construction & 0.00 & 0.01 & 0.01 & 0.01 \\
ind\_manufacture & 0.19 & 0.09 & 0.02 & 0.01 \\
ind\_retail & 0.17 & 0.14 & 0.03 & 0.02 \\
ind\_transport & 0.14 & 0.07 & 0.00 & 0.00 \\
ind\_information & 0.00 & 0.02 & 0.01 & 0.02 \\
ind\_finance & 0.43 & 0.17 & 0.02 & 0.01 \\
ind\_professional & 0.04 & 0.06 & 0.05 & 0.03 \\
ind\_education & 0.00 & 0.03 & 0.56 & 0.09 \\
ind\_leisure & 0.00 & 0.00 & 0.21 & 0.08 \\
ind\_services & 0.00 & 0.00 & 0.09 & 0.05 \\
ind\_public & 0.02 & 0.05 & 0.02 & 0.01 \\
educ\_lhs & 0.04 & 0.04 & 0.04 & 0.02 \\
educ\_hsg & 0.26 & 0.13 & 0.04 & 0.05 \\
educ\_sc & 0.50 & 0.15 & 0.12 & 0.06 \\
educ\_cg & 0.14 & 0.11 & 0.34 & 0.10 \\
educ\_ad & 0.07 & 0.07 & 0.46 & 0.08 \\
ms\_married & 1.00 & 0.02 & 0.02 & 0.04 \\
ms\_widowed & 0.00 & 0.00 & 0.03 & 0.07 \\
ms\_separated & 0.00 & 0.00 & 0.03 & 0.03 \\
ms\_divorced & 0.00 & 0.01 & 0.07 & 0.04 \\
ms\_nevermarried & 0.00 & 0.00 & 0.85 & 0.12 \\
region\_mw & 0.29 & 0.09 & 0.28 & 0.06 \\
region\_so & 0.47 & 0.11 & 0.25 & 0.06 \\
region\_we & 0.12 & 0.06 & 0.22 & 0.05 \\
region\_ne & 0.13 & 0.07 & 0.25 & 0.06 \\
\hline
\end{tabular}
\end{table}
We also test the statistical significance of the mean differences. Table \ref{table8} shows that the differences mentioned above are significant after controlling for simultaneous inference within categories, but only the differences in marital status, potential experience and education industry remain jontly significant at the 5\% level.
\begin{example}
Chardiff <- ca(fm = fmla1, data = wage2015, samp_weight = wage2015$weight, var = "male", 
               t = tw, cl = "diff", b = 500, cat = cat,  bc = FALSE, u = 0.05,
               subgroup = wage2015[, "female"] == 1,  boot_type = "weighted")
summary(Chardiff)
\end{example}
\begin{table}[ht]
	\centering
	\caption{Mean difference between 5\% most and least affected women}\label{table8}
	\begin{tabular}{lrrrr}
		\hline
		& Estimate & SE & JP-vals & Cat P-vals \\
		\hline
		lnw & -0.06 & 0.12 & 1.00 & 0.32 \\
		female & 0.00 & 0.00 &  &  \\
		exp1 & 18.12 & 3.49 & 0.09 & 0.00 \\
		occ\_manager & -0.55 & 0.18 & 0.49 & 0.03 \\
		occ\_service & -0.08 & 0.08 & 1.00 & 0.88 \\
		occ\_sales & 0.46 & 0.17 & 0.67 & 0.09 \\
		occ\_construction & -0.01 & 0.02 & 1.00 & 1.00 \\
		occ\_production & 0.18 & 0.09 & 0.89 & 0.24 \\
		ind\_minery & 0.00 & 0.02 & 1.00 & 1.00 \\
		ind\_construction & -0.01 & 0.02 & 1.00 & 1.00 \\
		ind\_manufacture & 0.18 & 0.10 & 0.97 & 0.79 \\
		ind\_retail & 0.14 & 0.14 & 1.00 & 0.99 \\
		ind\_transport & 0.14 & 0.07 & 0.90 & 0.63 \\
		ind\_information & -0.00 & 0.04 & 1.00 & 1.00 \\
		ind\_finance & 0.41 & 0.18 & 0.86 & 0.58 \\
		ind\_professional & -0.00 & 0.08 & 1.00 & 1.00 \\
		ind\_education & -0.56 & 0.10 & 0.07 & 0.04 \\
		ind\_leisure & -0.21 & 0.09 & 0.79 & 0.50 \\
		ind\_services & -0.09 & 0.05 & 0.98 & 0.82 \\
		ind\_public & -0.00 & 0.06 & 1.00 & 1.00 \\
		educ\_lhs & 0.00 & 0.05 & 1.00 & 1.00 \\
		educ\_hsg & 0.22 & 0.16 & 1.00 & 0.64 \\
		educ\_sc & 0.38 & 0.16 & 0.77 & 0.13 \\
		educ\_cg & -0.21 & 0.16 & 1.00 & 0.69 \\
		educ\_ad & -0.39 & 0.12 & 0.37 & 0.01 \\
		ms\_married & 0.98 & 0.05 & 0.00 & 0.00 \\
		ms\_widowed & -0.03 & 0.07 & 1.00 & 0.97 \\
		ms\_separated & -0.03 & 0.03 & 1.00 & 0.83 \\
		ms\_divorced & -0.07 & 0.05 & 0.99 & 0.56 \\
		ms\_nevermarried & -0.85 & 0.12 & 0.02 & 0.01 \\
		region\_mw & 0.01 & 0.15 & 1.00 & 1.00 \\
		region\_so & 0.22 & 0.17 & 1.00 & 0.50 \\
		region\_we & -0.10 & 0.11 & 1.00 & 0.75 \\
		region\_ne & -0.13 & 0.12 & 1.00 & 0.71 \\
		\hline
	\end{tabular}
\end{table}

Lastly we use show the functionality of the command \code{subpop}. We plot  projections of 90\% confidence sets for the 5\% most and least affected group with respect to two pairs of variables: log wages and potential experience, and marital status and potential experience. The estimated sets are obtained by weighted bootstrap with 500 repetitions and we drop the overlapped observations.
\begin{example}
set <- subpop(fm = fmla1, data = wage2015, var = "male", samp_weight = wage2015$weight, 
              boot_type = "weighted", b = 500, subgroup = wage2015[, "female"] == 1, 
              u = 0.05)

plot(set, varx = wage2015$exp1, vary = wage2015$lnw, main = "Projections of Exp-lnw", 
     sub = "OLS", xlab = "Exp", ylab = "Log Wages")

plot(set, varx = wage2015$exp1, vary =wage2015$ms, main = "Projections of Exp-MS", 
     sub = "OLS", xlab = "Exp", ylab = "Marital Status")
\end{example}
\begin{figure}[ht]
	\centering
	\begin{minipage}[b]{0.495\textwidth}
		\includegraphics[width=\textwidth]{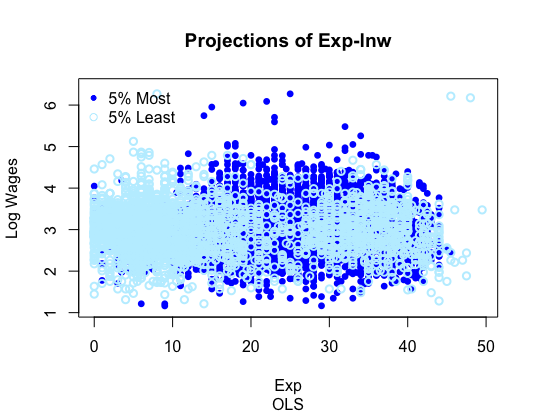}
	\end{minipage}
	\hfill
	\begin{minipage}[b]{0.495\textwidth}
		\includegraphics[width=\textwidth]{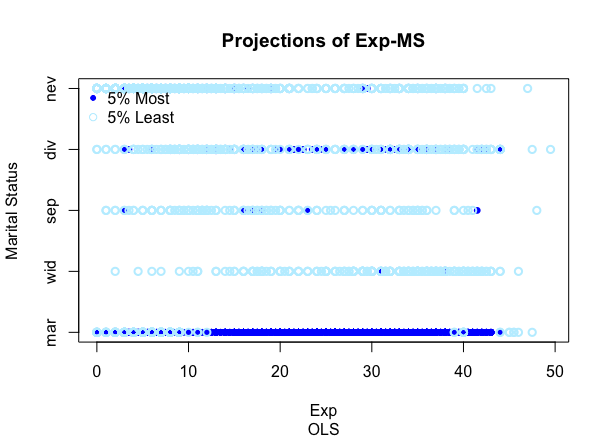}
	\end{minipage}
	\caption{Projections of confidence sets of 5\% most and least affected women.}\label{fig6}
\end{figure}

Figure \ref{fig6} shows that there are relatively more least affected women with low experience at all wage levels, more high affected women with high wages with between 15 and 45 years of experience, and more least affected women which are not married at all experience levels.
%%%%%%%%%%%%%%%%%%%%%%%%%%%%%%%%%%%%%%%%%%%%%%%%%%%%%%%%%%%%%%%%%%%%
\section{Acknowledgement}
We wish to thank the editor Norman Matloff, Thomas Leeper and an anonymous referee for insightful
comments that have helped improve the paper and package. We gratefully acknowledge research
support from the NSF.
%%%%%%%%%%%%%%%%%%%%%%%%%%%%%%%%%%%%%%%%%%%%%%%%%%%%%%%%%%%%%%%%%%%%
\bibliography{chen-chernozhukov-fernandezval-luo}

\address{Shuowen Chen\\
  Department of Economics\\
  Boston University\\
  270 Bay State Road, Boston, MA 02215\\
  USA\\
  \email{swchen@bu.edu}}

\address{Victor Chernozhukov\\
  Department of Economics\\
  Massachusetts Institute of Technology \\
  50 Memorial Drive, Cambridge, MA 02142 \\
  USA\\
  \email{vchern@mit.edu}}

\address{Iv{\'a}n Fern{\'a}ndez-Val\\
  Department of Economics\\
  Boston University\\
  270 Bay State Road, Boston, MA 02215\\
  USA\\
  \email{ivanf@bu.edu}}

\address{Ye Luo\\
  Faculty of Business and Economics\\
  The University of Hong Kong\\
  Cyberport Four, 100 Cyberport Rd, Telegraph Bay\\
  Hong Kong\\
  \email{hurtluo@hku.hk}}

\end{article}

\end{document}